\documentclass[letterpaper, amsfonts, amssymb, amsmath, reprint, showkeys, nofootinbib]{revtex4-1}

\usepackage{graphicx}
\usepackage{xcolor}
\usepackage{lipsum}
\usepackage{natbib}
\usepackage{physics}
\usepackage{vwcol}
\usepackage{hyperref}
\usepackage{upgreek}

\renewcommand{\mu}{\upmu}
\hypersetup{
    colorlinks = true,
    allcolors = black
}

\graphicspath{{figs/}}

\newcommand{\beginsupplement}{%
    \setcounter{table}{0}
    \renewcommand{\thetable}{S\arabic{table}}%
    \setcounter{figure}{0}
    \renewcommand{\thefigure}{S\arabic{figure}}%
    \renewcommand{\thesubsection}{\arabic{subsection}}
 }
 
\begin{document}
\title[Sutula et al]{Large-scale optical characterization of solid-state quantum emitters}

\author{Madison~Sutula$^1$}
\email{mmsutula@mit.edu}
\author{Ian~Christen$^1$}
\author{Eric~Bersin$^{1,2}$}
\author{Michael~P.~Walsh$^1$}
\author{Kevin~C.~Chen$^1$}
\author{Justin~Mallek$^2$}
\author{Alexander~Melville$^2$}
\author{Michael Titze$^3$}
\author{Edward~S.~Bielejec$^3$}
\author{Scott~Hamilton$^2$}
\author{Danielle~Braje$^2$}
\author{P.~Benjamin~Dixon$^2$}
\author{Dirk~R.~Englund$^1$}
\address{
$^1$Research Laboratory of Electronics, Massachusetts Institute of Technology, Cambridge, MA 02139, USA \\
$^2$Lincoln Laboratory, Massachusetts Institute of Technology, Lexington, MA 02421, USA\\
$^3$Sandia National Laboratories, Albuquerque, NM 87185, USA
}
\date{\today}

\begin{abstract}
Solid-state quantum emitters have emerged as a leading quantum memory for quantum networking applications. However, standard optical characterization techniques are neither efficient nor repeatable at scale. In this work, we introduce and demonstrate spectroscopic techniques that enable large-scale, automated characterization of color centers. We first demonstrate the ability to track color centers by registering them to a fabricated machine-readable global coordinate system, enabling systematic comparison of the same color center sites over many experiments. We then implement resonant photoluminescence excitation in a widefield cryogenic microscope to parallelize resonant spectroscopy, achieving two orders of magnitude speed-up over confocal microscopy. Finally, we demonstrate automated chip-scale characterization of color centers and devices at room temperature, imaging thousands of microscope fields of view. These tools will enable accelerated identification of useful quantum emitters at chip-scale, enabling advances in scaling up color center platforms for quantum information applications, materials science, and device design and characterization.
\end{abstract}


\maketitle

\begin{figure*}[htbp]
    \centering
    \includegraphics{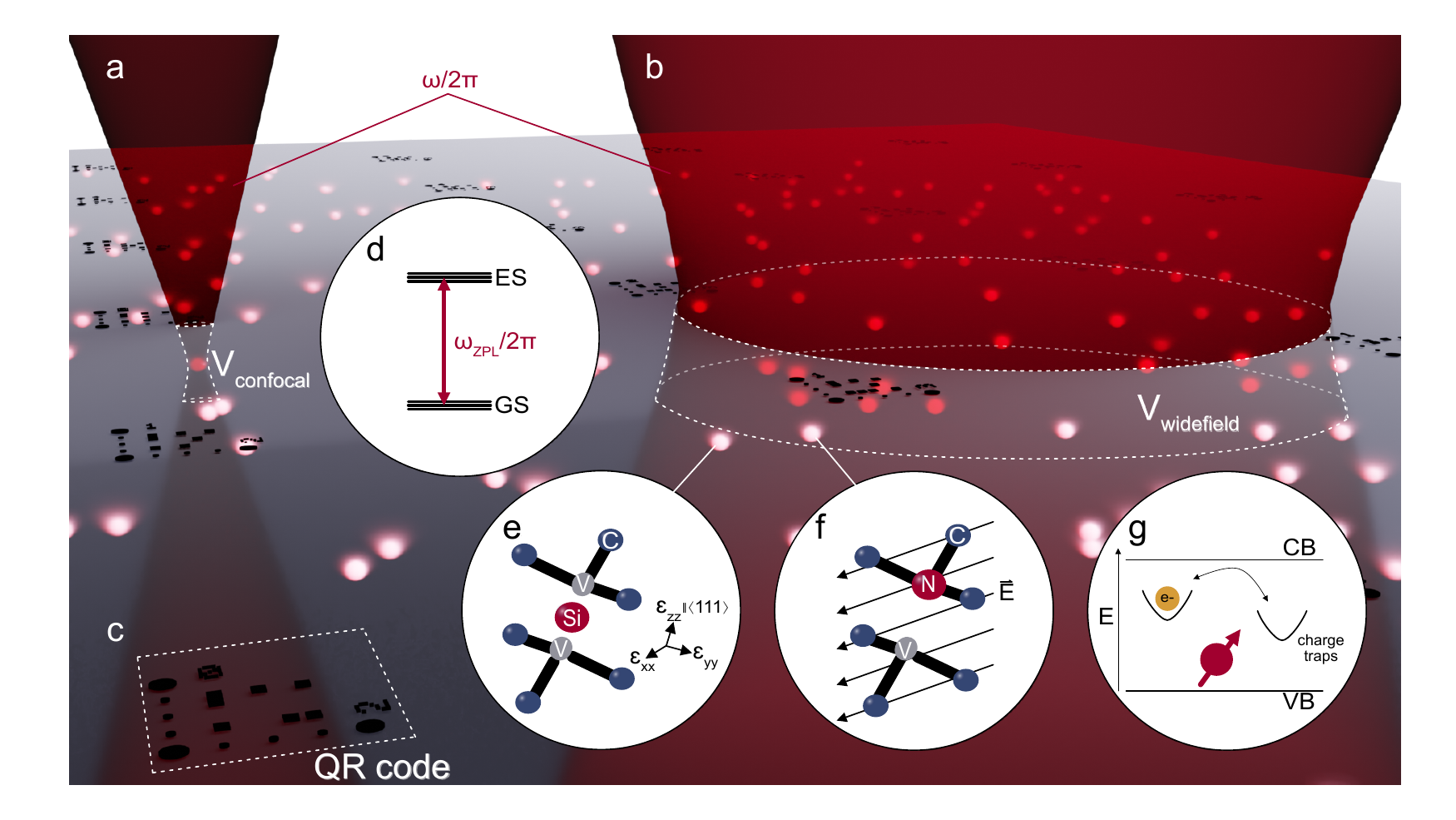}
    \caption{\textbf{Techniques for chip-scale characterization.} \textbf{a,} Standard confocal microscopy enables spectroscopy on a diffraction-limited volume, $\text{V}_{\text{confocal}}$. \textbf{b,} Our widefield microscopy technique discussed in Section~\ref{sec:wide} allows for parallelized spectroscopic measurements over many color centers simultaneously in a larger sample volume, $\text{V}_{\text{widefield}}$. \textbf{c,} We fabricate QR-style codes onto the surface of our samples as a spatial reference frame which we decode convolutionally in real time, enabling us to track and revisit color centers. Together, these techniques allow us to probe \textbf{d,} the optical transitions of color centers, including perturbations such as  \textbf{e,} local strain environment, which the silicon-vacancy center in diamond (pictured) is particularly sensitive to, \textbf{f,} electric fields, which the nitrogen-vacancy center in diamond (pictured) is particularly sensitive to, and \textbf{g,} the time-varying occupancy of charge traps in the crystal host, which is thought to cause inhomogeneous line broadening.}
    \label{fig:overview}
\end{figure*}

Quantum emitters play a central role in quantum information science and technology~\cite{Awschalom2021DevelopmentTechnologies,Alexeev2021QuantumDiscovery}. Color centers in solids have emerged as a leading platform for quantum information processing, with applications in sensing, computation, and communication. Their electron spin degree of freedom can store quantum states for milliseconds to seconds~\cite{Sukachev2017,Bar-Gill2013Solid-stateSecond}, and can be efficiently transduced into a flying qubit via a coherent spin-photon interface. These light-matter interactions can be engineered with cavity quantum electrodynamics~\cite{Nemoto2016PhotonicCenters,Janitz2020CavityDiamond}, accessible through nanofabrication~\cite{Siyushev2010MonolithicDetection,Hadden2010StronglyLenses,Mouradian2017RectangularDiamond,Calusine2014SiliconCenters,Nguyen2019AnDiamond,Rugar2021QuantumDiamond} and heterogeneous integration techniques~\cite{Bayat2014EfficientResonators,Mouradian2015ScalableCircuit,Janitz2015Fabry-PerotPhotonics, Gould2016LargeInformation,Wan2020Large-scaleCircuits}. Nuclear spin ancilla qubits provide additional degrees of freedom~\cite{Ruf2021QuantumDiamond}, enabling longer storage times~\cite{Bradley2019AMinute}, local multi-qubit logic protocols such as error detection and correction~\cite{Stas2022RobustDetection,Waldherr2014QuantumRegister}, and potential for more robust computation such as brokered entanglement~\cite{Benjamin2006BrokeredComputation} or cluster state generation~\cite{Choi2019}. 

In spite of these promising properties and early demonstrations,
most approaches rely on the integration and
control of a single color center at each node~\cite{Hermans2022QubitNetwork, Bhaskar2020,Pompili2021RealizationQubits}. However, quantum information processing nodes will require
many individually-addressable quantum emitters, each with long-lived spin states and a high-quality photonic interface~\cite{Lee2022ANetworks}. A critical challenge faced by color center qubits, especially when compared to neutral atom and trapped ion qubit architectures, is spectral inhomogeneity~\cite{Bersin2019IndividualVolume}. Although a number of techniques have been demonstrated to practically or theoretically overcome this inhomogeneity, including retuning emitter transitions via Stark~\cite{Bassett2011ElectricalFields} or strain tuning~\cite{Meesala2018StrainDiamond}, compensating for differential phase accumulation~\cite{Metz2008EffectProcessing}, and shifting emitted photons by optical frequency conversion~\cite{Stolk2022Telecom-BandCenters} or frequency modulation~\cite{Levonian2022OpticalEmitters}, each of these solutions relies on pre-characterization of the optical transitions involved. Additionally, up to now, approaches to understand the range of color center performance and uniformity have been limited to ensemble-level statistics, obscuring individual color center properties. Scaling up the number of solid-state quantum memories accessible in quantum information processing systems requires large-scale characterization techniques to identify the most viable qubits for practical use.

Here, we report on a high-throughput approach for spectroscopy and synthesis of quantum emitters, while retaining single-emitter resolution, on diamond color centers (Fig.~\ref{fig:overview}). We \hyperref[sec:tracking]{track color centers} by registering them to fabricated, machine-readable global coordinate system, enabling the systematic comparison of individual center sites over many experiments that is critical for understanding the role of materials processing~\cite{Walsh2020StatisticalDevices}. We then implement resonant photoluminescence excitation in a \hyperref[sec:wide]{widefield cryogenic microscope} to realize two orders of magnitude speed-up over confocal microscopy. Finally, we demonstrate \hyperref[sec:chipscale]{automated chip-scale characterization} of color centers and devices at room temperature, imaging thousands of microscope fields of view (see Extended Data Videos). 

With the large datasets accessible through these novel approaches, we identify a sample with an exceptionally narrow inhomogeneous distribution of silicon-vacancy centers in diamond (SiVs), and we verify in a sample implanted via focused ion beam (FIB) that the optical properties of highly strained SiVs do not degrade. These techniques provide a paradigm in which the optical properties of individual color centers can be identified, tracked, and categorized. We anticipate that the reported high-throughput tools will close the loop between materials processing and spectroscopy, enabling development of scalable quantum information processors. 

\section*{Results}

\section{Registering color centers to fabricated markers }\label{sec:tracking}

\begin{figure*}[htbp]
\centering
\includegraphics[width=\textwidth]{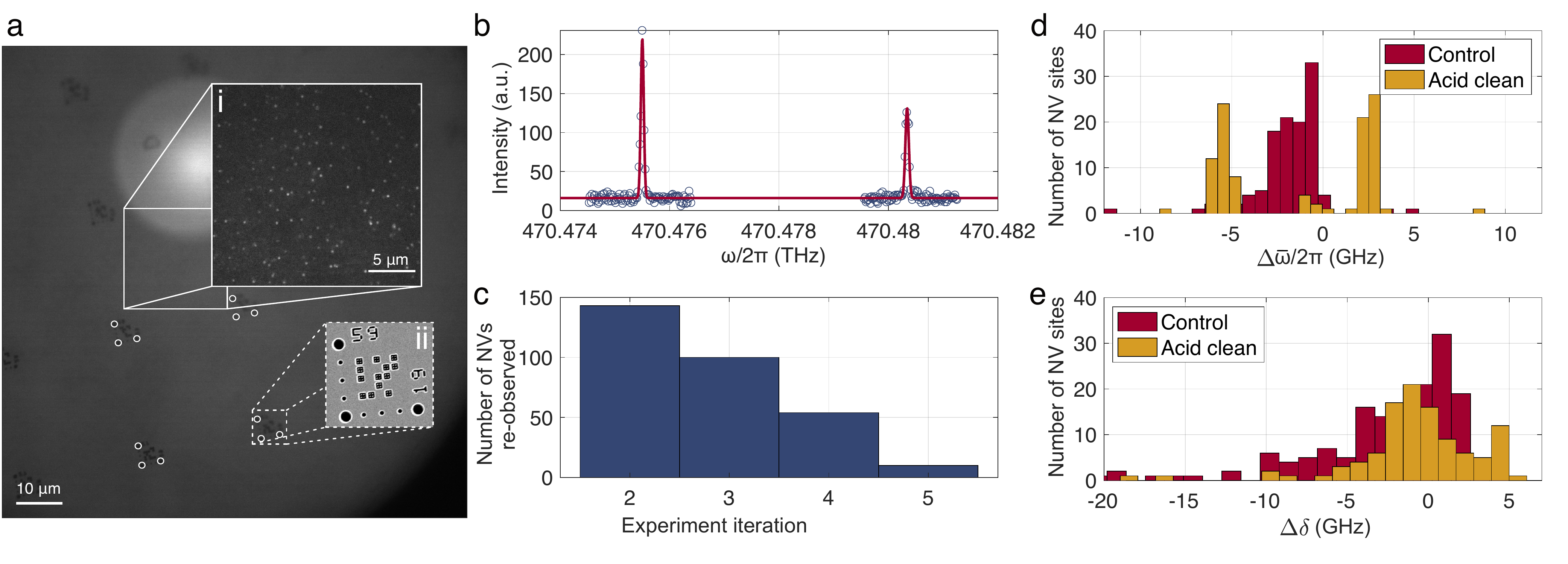}
\caption{ \textbf{Tracking color centers in a sample with fabricated QR codes.}
\textbf{a,} Confocal microscope image of a diamond patterned with QR codes and implanted with $^{15}$N. The full figure shows a confocal scan of the sample under incoherent illumination. The corners of four QR codes are emphasized with white circles. The insets show \textbf{i,} fluorescence from part of the full field of view under 515~nm excitation, with single nitrogen-vacancy centers in diamond (NVs), and \textbf{ii,} an SEM micrograph of a QR code. \textbf{b,} Typical photoluminescence excitation (PLE) spectrum for an NV center in this sample. \textbf{c,} Histogram of the number of experiments in which an NV was observed for a sample cooled down four times, with processing steps in between subsequent experiments. Errors in clustering sites are indicated in bin 5. \textbf{d,} Visualization of the shift in mean optical transition frequency $\Delta\bar{\omega}/2\pi = \bar{\omega}_i/2\pi-\bar{\omega}_j/2\pi$ of the two $\text{m}_\text{s}=0$ spin transitions between experiments $j=1$ \& $i=2$, which served as a control, and between experiments $j=1$ \& $i=3$, before and after tri-acid cleaning. After tri-acid cleaning, two populations of $\bar{\omega}/2\pi$ are observed, attributable to polarization due to the presence of an electric field caused by a change in surface termination. \textbf{e,} Visualization of the shift in the splitting $\Delta\delta = \delta_i - \delta_j$ between the two $\text{m}_\text{s}=0$ transitions between experiments $j=1$ \& $i=2$, which served as a control,  and between experiments $j=1$ \& $i=3$, before and after tri-acid cleaning.  }\label{fig:eg301}
\end{figure*}

Scanning confocal microscopy is a standard technique used to identify color centers, revealing bright spots located relative to the coordinates set by the scanning axes of the microscope. However, returning to a previously-characterized set of emitters on a chip can be challenging due to the vast difference between the $\mathcal{O}(100~\mu$m~$\times$~100~$\mu$m) size of the field of view and the $\mathcal{O}(5$~mm~$\times$~5~mm) size of the chip, especially after the frame of reference is lost to sample removal. Previous work has used bright `lodestar' emitters or image-stitching techniques to return to a given set of emitters~\cite{Chakravarthi2020WindowCenters}, though these techniques are not general, as not all samples possess sufficiently unique features for each field of view of interest.

We demonstrate the ability to register emitters to a global coordinate system using a standard home-built cryogenic confocal microscope setup. This technique enables us to revisit the same set of nitrogen-vacancy centers in diamond (NVs) over the course of multiple cooling cycles, and therefore to track emitter properties over time. We first fabricate quick response-style (QR) codes on the surface of the diamond chip (see Methods Section~\ref{methods:fab}). These are decoded in real-time with custom image processing tools by establishing a coordinate transform between the global sample coordinates and the local microscope coordinates using the region of sample space in view of the scanning confocal microscope (Fig.~\ref{fig:eg301}a). 

We use off-resonant scanning fluorescence microscopy under 515~nm excitation to image a low-density of NV centers on the surface of the chip. To create these fluorescent centers, the chip was implanted with $1\times 10^9/\text{cm}^2$ of $^{15}$N ions at 185~keV, and subsequently annealed at 1200$^\circ$C in ultra-high vacuum, tri-acid cleaned in a 1:1:1 mixture of nitric, perchloric, and sulfuric acids by boiling for 1 hour~\cite{Chu2014CoherentCenters}, and finally cleaned in a 3:1 sulfuric~acid:hydrogen~peroxide piranha solution.
We classify bright, diffraction-limited spots as candidate emitter sites, as shown in the inset of Fig.~\ref{fig:eg301}a. We then determine the spectral positions and linewidths of the zero-phonon line (ZPL) optical transitions at each candidate site with photoluminescence excitation (PLE) spectroscopy. To do this, we measure the fluorescence response of each emitter by scanning a resonant laser over the ZPL transitions and collecting light emitted into the phonon side-band (PSB) on an avalanche photodiode. We label the spin conserving transitions associated with $\text{m}_\text{s}=0$ by mean ZPL frequency $\bar{\omega} = 1/2 \left(E_x+E_y\right)$ and splitting $\delta = |E_y-E_x|$~\cite{Bassett2011ElectricalFields}. Further information about the experimental setup and NV centers is provided in the Supplementary Information. A typical NV PLE spectrum is shown in Fig.~\ref{fig:eg301}b. Details regarding data analysis and QR encoding are provided in the Methods Sections~\ref{methods:confocal} and \ref{methods:qr}.

We track NV centers over four experiments by registering emitters to the sample rather than to the microscope coordinates, in spite of the stochastic movement of the sample caused by remounting along with warming and cooling the system between experiments. Experiments 1 and 2 served as a control, as we thermally cycled and remounted the sample in between, but performed no additional materials processing. Between experiments 2 and 3, the sample was tri-acid cleaned in a 1:1:1 mixture of nitric, perchloric, and sulfuric acids by boiling for 1 hour~\cite{Chu2014CoherentCenters} and between experiments 3 and 4, the sample was annealed in an oxygen environment at 450$^\circ$C for 4~hours~\cite{Fu2010ConversionOxidation}. In each experiment, NVs and QR codes were registered to the local microscope coordinate system, and then transformed to the global sample coordinate system.  After registering the locations of NVs in all four experiments, we identified the NVs in each experiment that belonged to the same confocal site on the sample using a clustering algorithm constrained by a diffraction-limited Euclidean threshold distance. The number of experiments in which a color center was found at the same global sample coordinate site is provided in Fig.~\ref{fig:eg301}c: we observe that some NVs disappear over multiple experiments.

By tracking individual color centers over multiple cryostat cooling cycles, we observe that tri-acid cleaning the diamond leads to shifts in the mean frequency at a given NV site $\Delta\bar{\omega}/2\pi = \bar{\omega}_i/2\pi - \bar{\omega}_j/2\pi$ (for experiments $i,j$) of the $\text{m}_\text{s}=0$ optical transitions exceeding those observed in the control. Conversely, the change in splitting $\Delta\delta = \delta_i - \delta_j$ between the two $\text{m}_\text{s}=0$ states does not deviate from the control. These observations, as shown in Fig.~\ref{fig:eg301}d-e, are consistent with the addition of an electric field normal to the surface~\cite{Bassett2011ElectricalFields}, attributable to a changed surface termination~\cite{Sangtawesin2019OriginsSpectroscopy}. Notably, the population-level statistics for $\bar{\omega}/2\pi$ alone do not reveal this phenomenon; tracking individual emitters uniquely enabled us to observe these minute spectral shifts.
\begin{figure*}[tbp]
\centering
\includegraphics[width=\textwidth]{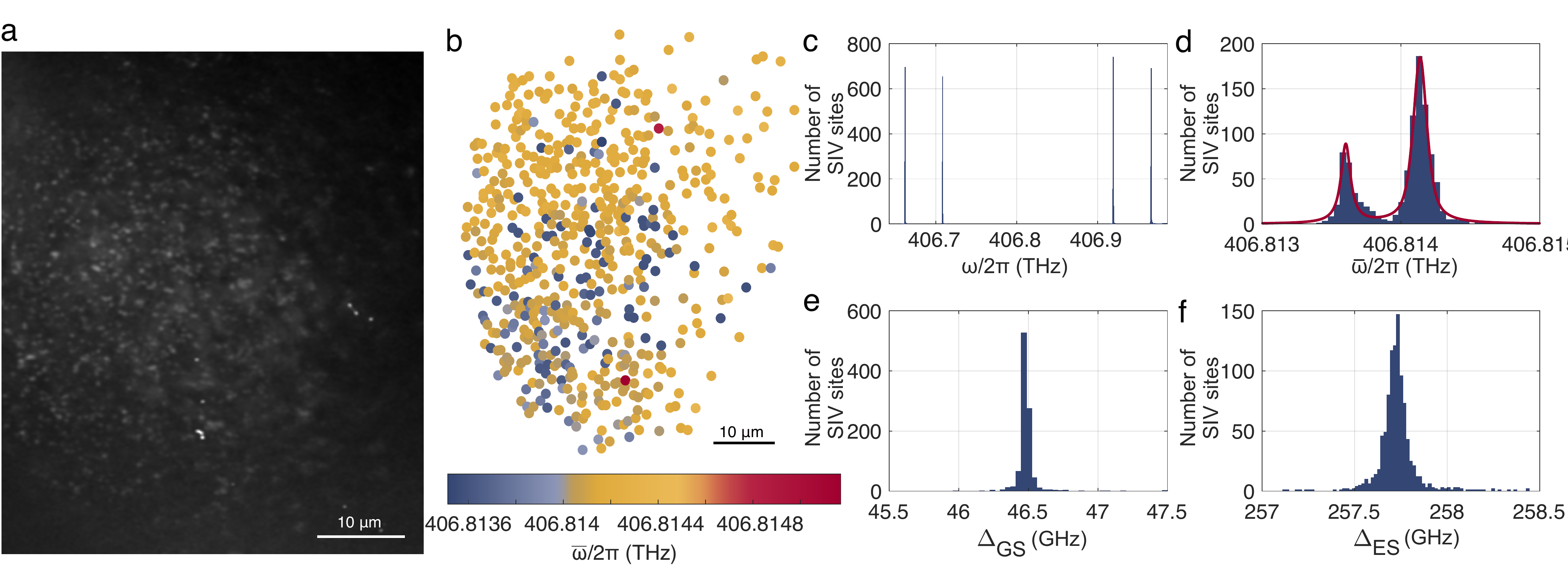}
\caption{\textbf{Widefield photoluminescence excitation of silicon-vacancy centers.}
\textbf{a,} Fluorescence image of Sample A reconstructed from widefield PLE as the resonant laser frequency $\omega/2\pi$ was swept over 10~GHz, using the same field of view as~\cite{Christen2022AnControl}. \textbf{b,} Strain map indicating the spatial distribution of the mean zero-phonon line (ZPL) transition $\bar{\omega}/2\pi$ (a proxy for axial strain) at 584 sites that had one PLE peak at each of the four SiV transition frequencies. We observe that the SiVs are split into two classes of axial strain. \textbf{c,} Histograms of the locations of the four optical transitions for 984 groups of SiV peaks, where each group of SiV peaks consists of a peak at each of the four SiV transition frequencies. \textbf{d,} Mean ZPL transition frequency $\bar{\omega}/2\pi$ for 984 groups of SiV peaks: the split into two classes of axial strain is apparent in the distribution. Fitting the empirical probability density function ($\sigma=10$~MHz, see Methods Section~\ref{methods:pdf}) shown in red reveals a population centered at 406.8141 THz with standard deviation 59~MHz, and a population centered at 406.8136~THz with standard deviation 48~MHz. \textbf{e,} Ground state splitting ($\Delta_{GS}$), and \textbf{f,} Excited state splitting ($\Delta_{ES}$), for 984 groups of SiV peaks. The ground and excited state splittings closely approximate the splitting due to spin-orbit coupling alone, revealing the low-strain environment. }\label{fig:LE7}
\end{figure*}

\begin{figure*}
\centering
\includegraphics[width=\textwidth]{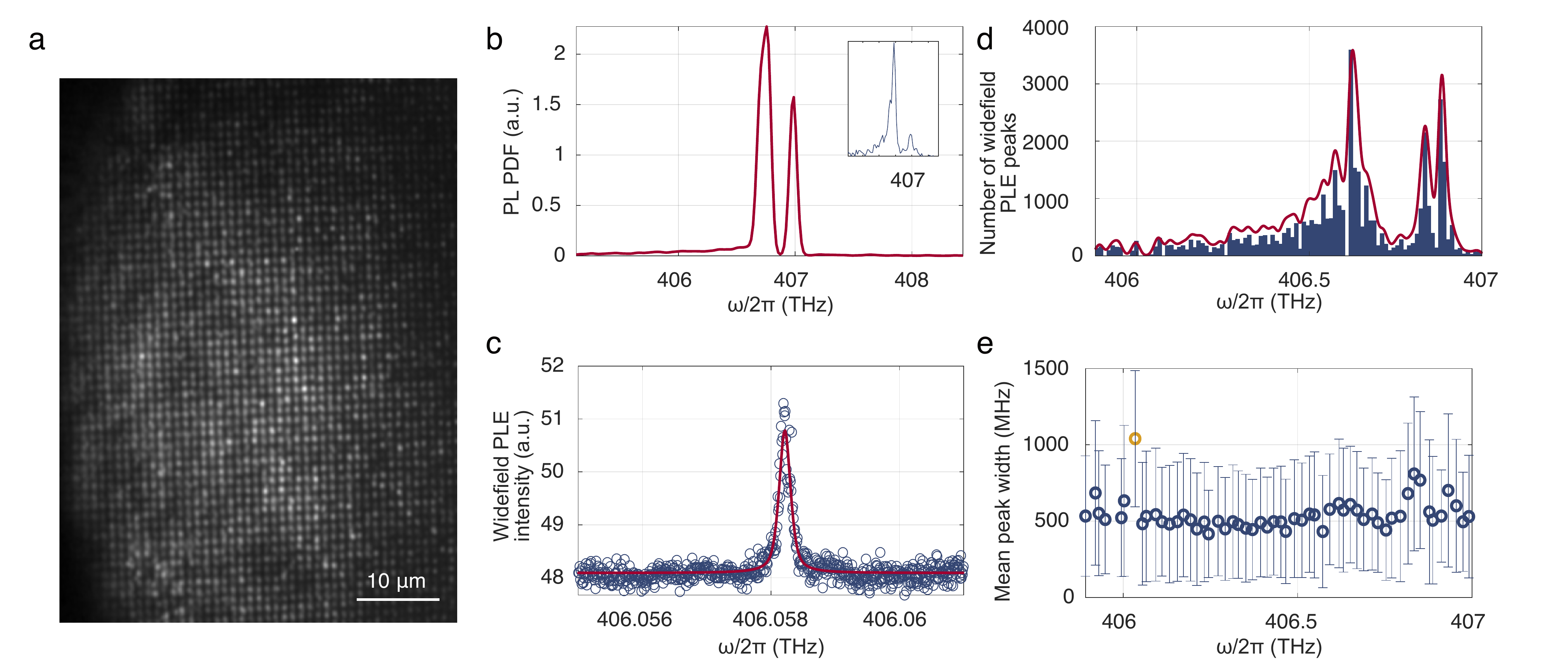}
\caption{ \textbf{Widefield photoluminescence excitation in a FIB-implanted sample.}
\textbf{a,} Fluorescence image of Sample B reconstructed from widefield PLE as the resonant laser frequency $\omega/2\pi$ was swept over 10~GHz. The grid of focused ion beam implantation sites is clearly visible. \textbf{b,} Probability density function showing the probability to find a photoluminescence (PL) peak as a function of wavelength for 11,010 SiVs measured confocally under 515~nm excitation and collected on a spectrometer. The inset shows the PL spectrum at one site. \textbf{c,} Typical widefield PLE spectrum of one SiV transition at one spatial site in the widefield field of view, with intensity data (blue circles) fit to a pseudo-Voigt function (red line). The fit center location is 406.0582~THz, and the fit linewidth is 210.6(1)~MHz. Details of the image processing performed to extract widefield PLE spectra can be found in Methods Section~\ref{methods:wf}.  \textbf{d,} Histogram showing inhomogeneous distribution of SiV PLE measured in widefield as a function of resonant laser frequency, $\omega/2\pi$. 40,186 peaks with linewidths less than 2~GHz and greater than the lifetime limit of $(2\pi\times1.7~\text{ns})^{-1}$ are represented. A kernel estimate of the empirical probability density function (see Methods Section~\ref{methods:pdf}) is shown in red. \textbf{e,} Distribution of linewidths for the 40,186 peaks shown in \textbf{d}. The mean linewidth of all peaks that line in each 20~GHz bin is shown as a blue circle. Bins containing fewer than 10 peaks are shown in yellow, and discarded from analysis. Error bars give the standard deviation of the mean. Optical linewidths do not significantly increase at lower frequencies, indicating that the optical properties of highly-strained emitters do not degrade. }\label{fig:UC2}
\end{figure*}

\begin{figure*}
\centering
\includegraphics[width=\textwidth]{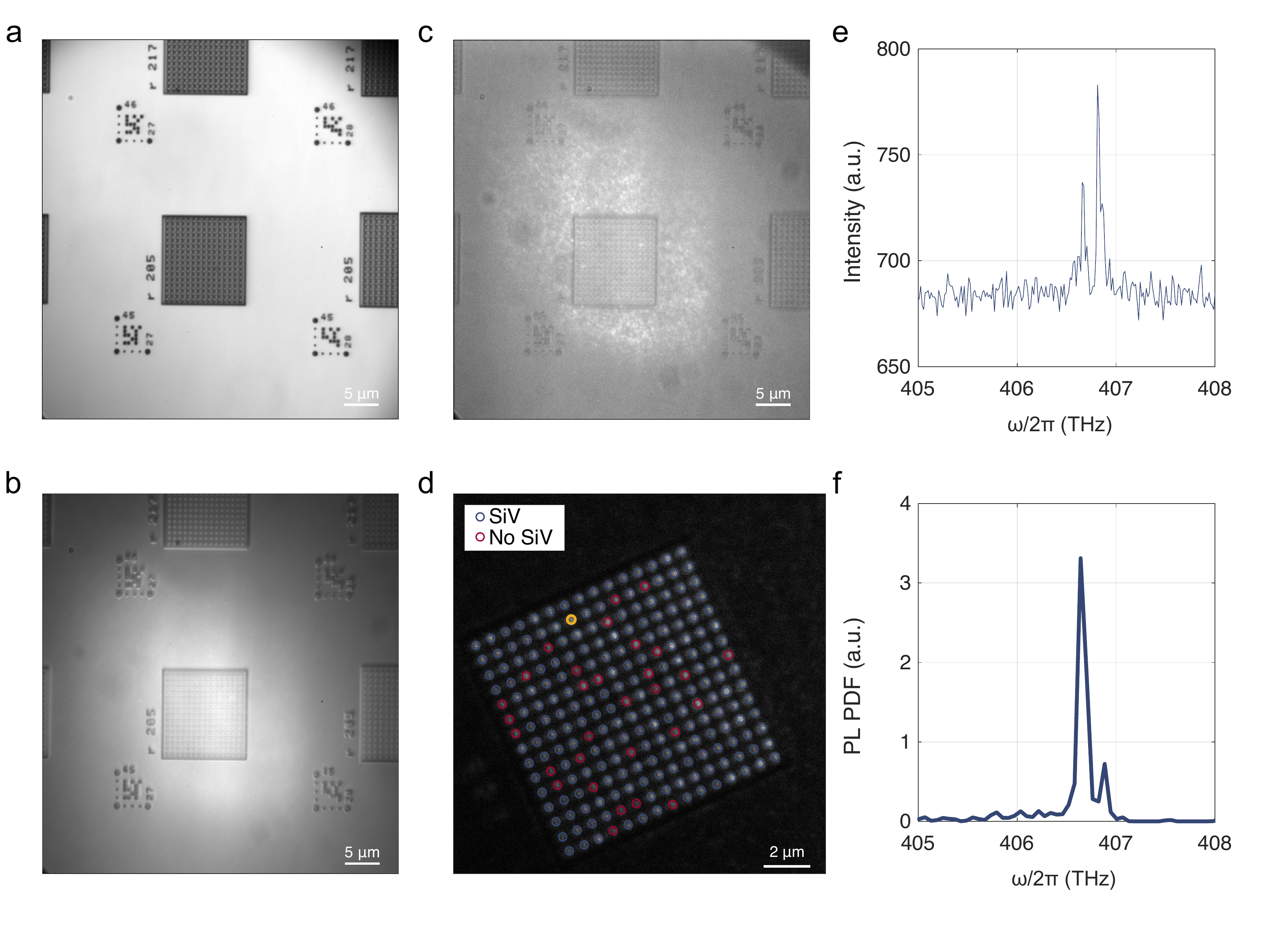}
\caption{\textbf{Chip-scale verification of fabricated devices at room temperature.} \textbf{a,} White light image of one QR region. \textbf{b,} Photoluminescence (PL) from the region shown in \textbf{a} under 532~nm excitation with a longpass filter to block the excitation laser. \textbf{c,}  PL from the region shown in \textbf{a} under 532~nm excitation with a bandpass filter to highlight SiV PL. A video demonstrating the automated imaging process in action over many QR regions is provided as Extended Data Video 2. \textbf{d,} Following chip scale characterization, promising regions are investigated at cryogenic temperatures. Here, a confocal image of the region shown in \textbf{a} (rotated 30 degrees counter-clockwise) is overlaid with green circles to show pillars that contained SiVs, and blue circles to show pillars that did not contain SiVs. \textbf{e,} Example PL spectrum at cryogenic temperatures from the pillar circled in yellow in \textbf{d}. \textbf{f,} Probability density function showing the probability to find a PL peak as a function of wavelength for all pillars with SiVs shown in \textbf{d}.}\label{fig:ROMI}
\end{figure*}
\section{Parallelized photoluminescence excitation via cryogenic widefield microscopy}\label{sec:wide}

Although the ability to register the positions of single emitters to the surface of a sample enables individual color centers to be tracked and compared over time, the characterization rate for the previous experiments is limited by the time to obtain one resonant PLE scan, as each site must be measured sequentially. To overcome this bottleneck, we modify our standard confocal microscope setup to excite an entire field of view with a tunable resonant laser in widefield mode (see Supplementary Information), and thus parallelize PLE measurements to realize a dramatic speed-up. This is made possible by the introduction of an electron-multiplying charge coupled device (EMCCD) camera in the imaging path, which enables us to resolve individual diffraction-limited color centers via image processing, as detailed in Methods Section~\ref{methods:wf}. The largest speedup that can be achieved with this techniques is set by the number of diffraction-limited emitter sites in a given field of view; we provide a detailed discussion of the regimes in which our widefield technique is faster than confocal microscopy in Methods Section~\ref{methods:speed}. We demonstrate widefield PLE for two systems containing silicon-vacancy centers in diamond (SiVs): one with an exceptionally narrow inhomogeneous distribution of optical transitions (Sample A), and one with a broad distribution (Sample B). 

For Sample A, SiVs were incorporated in-situ during chemical vapor deposition (CVD) overgrowth on a low-strain substrate. The sample was subsequently tri-acid cleaned and annealed at 1200$^\circ$C in ultra-high vacuum. Fig.~\ref{fig:LE7}a shows a fluorescence map reconstructed from widefield PLE measurements over the SiV zero-phonon line (ZPL) (specifically using the ground spin-orbit conserving C transition). By measuring PLE over all four SiV ZPL transitions (see Extended Data Video 1), we can determine the strain environment in the overgrown layer throughout the entire field of view, revealing two classes as shown in Fig.~\ref{fig:LE7}b-d. We ascribe this to two populations of strain along the axis of the SiV. We generate an empirical probability distribution (PDF) function ($\sigma =$ 10~MHz, see Methods Section~\ref{methods:pdf}) to determine the inhomogeneous distribution. A Gaussian fit to the empirical PDF representing $\bar{\omega}/2\pi$ reveals a population centered at 406.8141 THz with standard deviation 59 MHz, and a population centered at 406.8136 THz with standard deviation 48 MHz. Previous work has demonstrated that CVD growth is a promising path towards generating a narrow inhomogeneous distribution of SiVs~\cite{Rogers2014All-opticalDiamond}, but to our knowledge, our measurement represents the largest number of SiVs with center frequencies that lie within the bandwidth given by the lifetime-limited linewidth, enabling work requiring high spectral homogeneity~\cite{Christen2022AnControl}. In a single field of view, we measure 257 SiV sites with ground spin-orbit C-transition within $(2\pi\times1.7~\text{ns})^{-1}$ from 406.7080~THz, and 613 SiV sites with ground spin-orbit C-transition within $(2\pi\times1.7~\text{ns})^{-1}$ from 406.7085~THz. The low-transverse-strain environment is evident in the ground state ($\Delta_{GS}$) and excited state ($\Delta_{ES}$) splittings, which closely approximate the splittings due to spin-orbit coupling~\cite{Meesala2018StrainDiamond} (Fig.~\ref{fig:LE7}e-f). Further information about the relationship between SiV optical transitions and strain environment can be found in the Supplementary Information. Given the emitter density and inhomogeneous distribution in this field of view, our widefield technique yields data on the strain environment $>700$ times faster than standard confocal microscopy (see Methods Section~\ref{methods:speed}), saturating the limit of achievable speed-ups given the emitter density in the sample.

Having demonstrated the potential for widefield PLE to generate comprehensive datasets more quickly than is possible with standard confocal PLE, we then perform widefield PLE on Sample B, where SiVs were incorporated by focused-ion beam (FIB) implantation of $^{29}$Si and subsequent high-temperature annealing (see Methods Section~\ref{methods:FIB}). Individual FIB sites are visible in a reconstructed PLE fluorescence map covering a fraction of the $>$1~THz inhomogeneous distribution of all ZPL transitions, as shown in Fig.~\ref{fig:UC2}a.

A typical widefield PLE spectrum at one site is shown in Fig.~\ref{fig:UC2}c. After fitting the widefield PLE spectra at all candidate sites, we extract the locations of 40,186 widefield PLE peaks, as shown in Fig.~\ref{fig:UC2}d, as well as the linewidths.  We measure widefield PLE over the entire inhomogeneous distribution, illuminating a long tail of highly-strained $^{29}$Si present in the FIB-implanted sites. We validate the presence of the long tail of highly strained SiVs using off-resonant PL under 515~nm illumination, as shown in Fig.~\ref{fig:UC2}b. To better understand the role that strain plays in SiV optical linewidths, we section the data into 20~GHz bins, and generate an empirical probability density function (see Methods Section~\ref{methods:pdf}) of the optical linewidths of all peaks in each bin. We fit this data to a Gaussian distribution, and plot the mean linewidth in each frequency bin in Fig.~\ref{fig:UC2}e. The error bars represent one standard deviation from the mean. We find that optical linewidths do not significantly degrade at low optical frequencies, which is an important validation for high-temperature operation of SiV quantum memories using highly-strained emitters~\cite{Stas2022RobustDetection}.

Widefield PLE enables us to collect data for every diffraction-limited site within the field of view of our camera and within the range of our spectral scan, reducing pre-selection bias. This permits a more honest and comprehensive view into color center performance, rather than measuring color centers serially until an emitter with the requisite optical properties for a given application is found. Beyond gaining statistics on the likelihood of finding a high quality quantum emitter, this technique enables large-scale characterization of materials properties in the environment surrounding color centers, for example as demonstrated here by measuring strain.

\section{Chip-scale verification of fabricated devices at room temperature}\label{sec:chipscale}

In addition to measuring quantum emitter properties, it is also critical to be able to characterize color center generation. Further, thousands of photonic devices can be fabricated on a single chip; characterizing the properties and promise of each of these devices is time-consuming and tedious. We automate widefield imaging and spectroscopy across a $2\times2$~mm region of a diamond chip, stepping serially between QR regions and using fluorescence to qualitatively observe color centers in nanopillar devices within each field of view (see Extended Data Video 2). A room-temperature setup with a motorized sample stage and home-built imaging processing software enables us to move from region to region, using real-time convolutional QR code detection as feedback to bring the sample into focus and compensate for variations in the sample stage motion (see Methods Section~\ref{methods:qr}). Importantly, this process, and our QR detection technique, is robust against the presence of dirt or complex fabricated structures (see Extended Data Videos 3 and 4). Fig.~\ref{fig:ROMI} shows a single QR region under incoherent reflection imaging for focusing and registration, as well as under fluorescence imaging with a 532~nm excitation laser. This set of nanopillars was identified to have bright pillars at room temperature. To confirm this technique enables identification of color centers that have been successfully incorporated into nanostructures, we verify that the pillars contain SiVs by measuring confocal PL under 515~nm excitation at 4~K. 

We regularly gather datasets over samples with $\mathcal{O}(10~\text{mm}^2)$ area, corresponding to $\mathcal{O}(10^8)$ diffraction-limited spots. Moreover, the addition of an automated filter wheel allows us to coarsely measure the spectral content of each field of view: Extended Data Video 5 displays data acquired with 10~nm bandpass filters around 737~nm, 620~nm, and 600~nm targeting silicon-, tin-, and germanium-vacancy color centers on red, green, and blue color channels, respectively. This imaging enables chip-scale characterization and classification to identify promising devices and color centers in a time-efficient and automated manner.

\section*{Discussion}
Large-scale characterization of color centers and photonic devices is necessary to scale solid-state devices from proof-of-principle demonstrations to useful technological applications. As the number of devices on chip increases to thousands or more, manual point-by-point characterization becomes intractable; our suite of automated tools performs spectroscopic measurements hundreds of times faster than traditional cryogenic confocal microscopy, opening up a new regime of system scaling. Individual color center tracking reveals materials-level insight into device performance which can be used to inform and revise qubit fabrication and processing. The resultant feedback loop between design, fabrication, and characterization offers a pathway to improved yields and device performance. 

Furthermore, characterization of completed devices is critical in assessing device properties and performance before integration into larger systems, a process realized at chip-scale by our system. We emphasize that these techniques are readily extensible to other materials platforms and quantum emitters~\cite{Viitaniemi2022CoherentNanowires,Higginbottom2022OpticalSilicon,Lukin20204H-silicon-carbide-on-insulatorPhotonics,Dietz2022Spin-AcousticCarbide,Babin2022FabricationCoherence,Hu2020PhotoluminescenceExcitation,Trusheim2020Transform-LimitedDiamond,Davidsson2022Exhaustive4H-SiC}, providing a basis for broad implementation.

\appendix
\beginsupplement

\pagebreak
\section*{Methods}
\subsection{Empirical determination of probability density functions}\label{methods:pdf}
We estimate empirical probability density functions using the kernel distribution given by \\
\begin{equation}
    f_k(\lambda) = \frac{1}{n} \sum_{i=1}^n K\left(\frac{\lambda-\lambda_i}{h} \right)
\end{equation}
where $n$ is the number of samples, $\lambda_i$ are the sample values, $K$ is the smoothing function, and $h$ is the smoothing bandwidth. We used a Gaussian smoothing function with bandwidth $\sigma$: 
\begin{equation}
    K(\lambda) = \text{exp}\left(\frac{\left(\lambda - \lambda_c\right)^2}{2\sigma^2}\right)
\end{equation}
Unless otherwise noted, the bandwidth $\sigma$ is given by our spectrometer limited bandwidth of 0.05 nm (for photoluminescence) or the appropriate lifetime limited linewidth (for photoluminescence excitation), and $\lambda_c$ for each candidate site was set to be the center of the fit spectra at each candidate site.

\subsection{Confocal data analysis}\label{methods:confocal}
\begin{figure}[htbp]
\centering
\includegraphics[width=\linewidth]{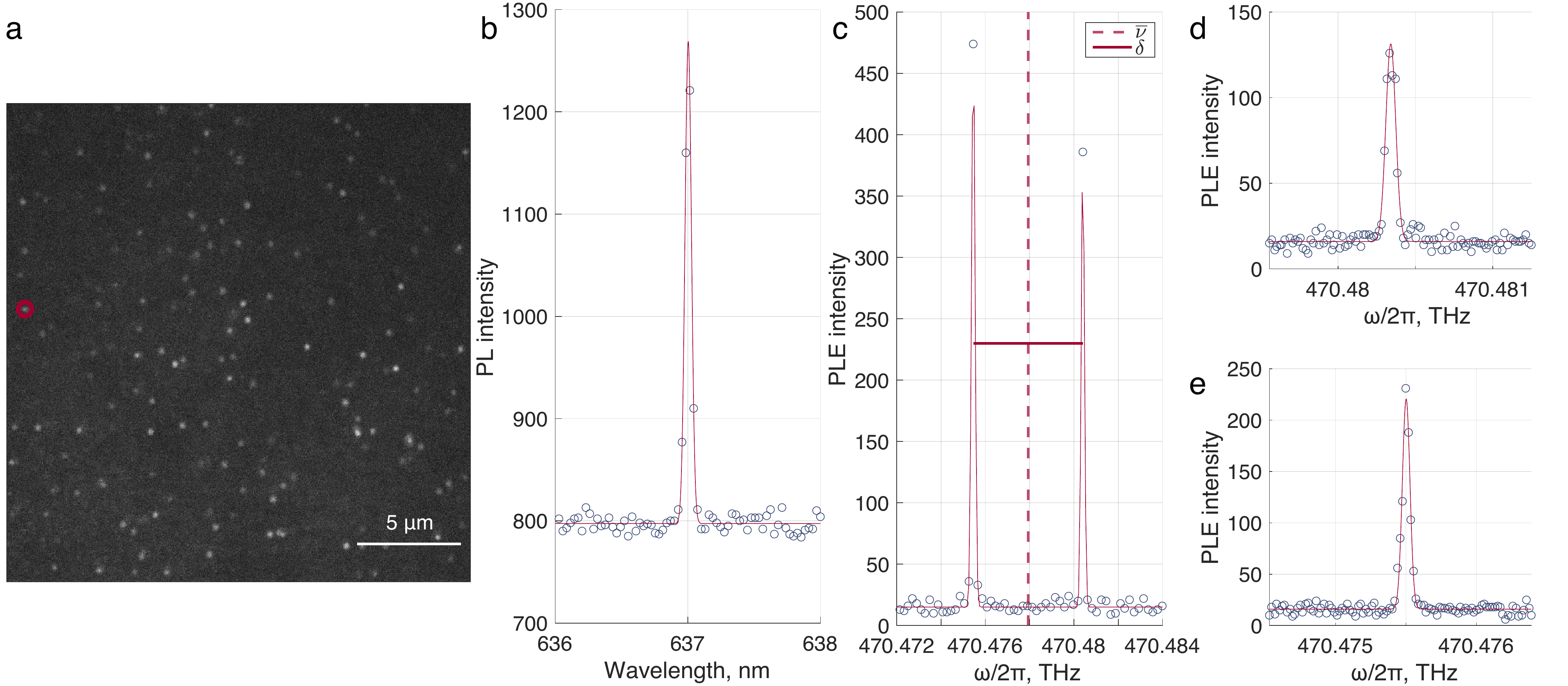}
\caption{\textbf{Experimental stack for confocal spectroscopy.} \textbf{a,} Confocal fluorescence image of nitrogen-vacancy centers in diamond under 515~nm excitation. \textbf{b,} Confocal photoluminescence spectrum under 515~nm excitation of the NV site highlighted in red in \textbf{a}. Only one peak is resolvable on the spectrometer: the center of the fit to this peak is used to center the resonant laser for resonant spectroscopy. \textbf{c,} Open-loop resonant photoluminescence excitation spectrum of the NV site, showing the $E_x$ and $E_y$ transitions. The center location and  $E_x$-$E_y$ splitting, as well as the fit used to determine these values, are overlaid. \textbf{d,}-\textbf{e,} Closed-loop photoluminescence excitation scans of the spectra shown in \textbf{c}, with fits overlaid. }
\label{fig:confocal}
\end{figure}

To identify and characterize color centers, we first measure a galvanometer scan of photoluminescence excited with a strong 515~nm excitation laser (Fig.~\ref{fig:confocal}a). We collect fluorescence on an avalanche photodiode in our cryogenic confocal microscope, which is described in detail in the Supplementary Information. We identify candidate sites by first applying a Gaussian bandpass filter to the aggregate image $A(x,y)$, and then fit 2D Gaussians to the resulting image. First, we convolve the aggregate image with a 2D Gaussian with the lowpass standard deviation $\sigma_{LP}$:
\begin{equation}
f_{LP}(x,y) = e^{-\frac{\left(x^2+y^2\right)}{2\sigma_{LP}^2}}
\label{eq:lowpass}
\end{equation}
We then discretize and convolve this kernel with the aggregate image $I(x,y)$ via discrete 2-dimensional convolution to yield an intermediate filtered image $R(x,y)$:
\begin{equation}
    R(x,y) = \sum_i \sum_j I(i,j) f_{LP}(x-i+1,y-j+1)
    \label{eq:filtered}
\end{equation}
We high-pass filter the resulting image $R(x,y)$ by convolution with a 2D Gaussian with high pass standard deviation $\sigma_{HP}$:
\begin{equation}
f_{HP}(x,y) = e^{-\frac{\left(x^2+y^2\right)}{2\sigma_{HP}^2}}
    \label{eq:highpass}
\end{equation}
Finally, we discretize and convolve this kernel with the intermediate image $R(x,y)$ via discrete 2-dimensional convolution to yield a bandpass-filtered image $F(x,y)$:
\begin{equation}
    F(x,y) = R(x,y)- \sum_i \sum_j R(i,j) f_{HP}(x-i+1,y-j+1)
    \label{eq:bandpassfilteredim}
\end{equation}\\
We use this Gaussian bandpass filtering process to identify bright spots with radii between the lowpass and highpass filter standard deviations. By fitting 2D Gaussians to the filtered image, we generate candidate emitter sites, which are stored as an array of galvanometer coordinates.

We then revisit each candidate site to determine its spectral position: we collect photoluminescence excited with a strong 515~nm laser onto a spectrometer, as shown in Fig.~\ref{fig:confocal}b. We fit the spectrum taken at each candidate site and determine the center zero-phonon line (ZPL) frequency up to the resolution limit of the spectrometer. Once the spatial and coarse spectral coordinates are identified, we measure an open-loop photoluminescence excitation spectrum at each candidate site using a Newport Velocity TLB-6704 laser, as shown in Fig.~\ref{fig:confocal}c. We repump the emitters in between each frequency step to initialize the NVs into the ground $\text{m}_\text{s}=0$ state, enabling us to probe the $E_x$ and $E_y$ spin conserving transitions associated with $\text{m}_\text{s}=0$. The mean frequency of the ZPL transition $\bar{\nu}=\bar{\omega}/2\pi$ and splitting $\delta$ associated with the emitter at this site are indicated in Fig.~\ref{fig:confocal}c. Finally, we obtain a closed-loop high resolution scan, in which we feed back on the laser frequency using a HighFinesse WS7 wavemeter, of each transition at every candidate site, as shown in Fig.~\ref{fig:confocal}d-e. 

\subsection{Widefield data analysis}\label{methods:wf}
To measure widefield PLE in Samples A and B, we swept the frequency of a ti:sapph excitation laser (M$^2$ SolsTiS) over the inhomogeneous distribution in frequency steps much smaller than the emitter lifetime, using the setup described in the Supplementary Information. For Sample B, weak 515~nm Cobolt laser was applied continuously to repump the emitters. We tuned the resonator internal to the resonant laser in voltage steps that corresponded to 10 MHz steps in frequency, and locked the laser to a HighFinesse WS7 Wavemeter (60 MHz absolute accuracy, 2 MHz resolution) at each step. Once tuning was complete, the EMCCD was exposed for 0.5~s for Sample A and for 1~s for Sample B.

\begin{figure}[htbp]
\centering
\includegraphics[width=\linewidth]{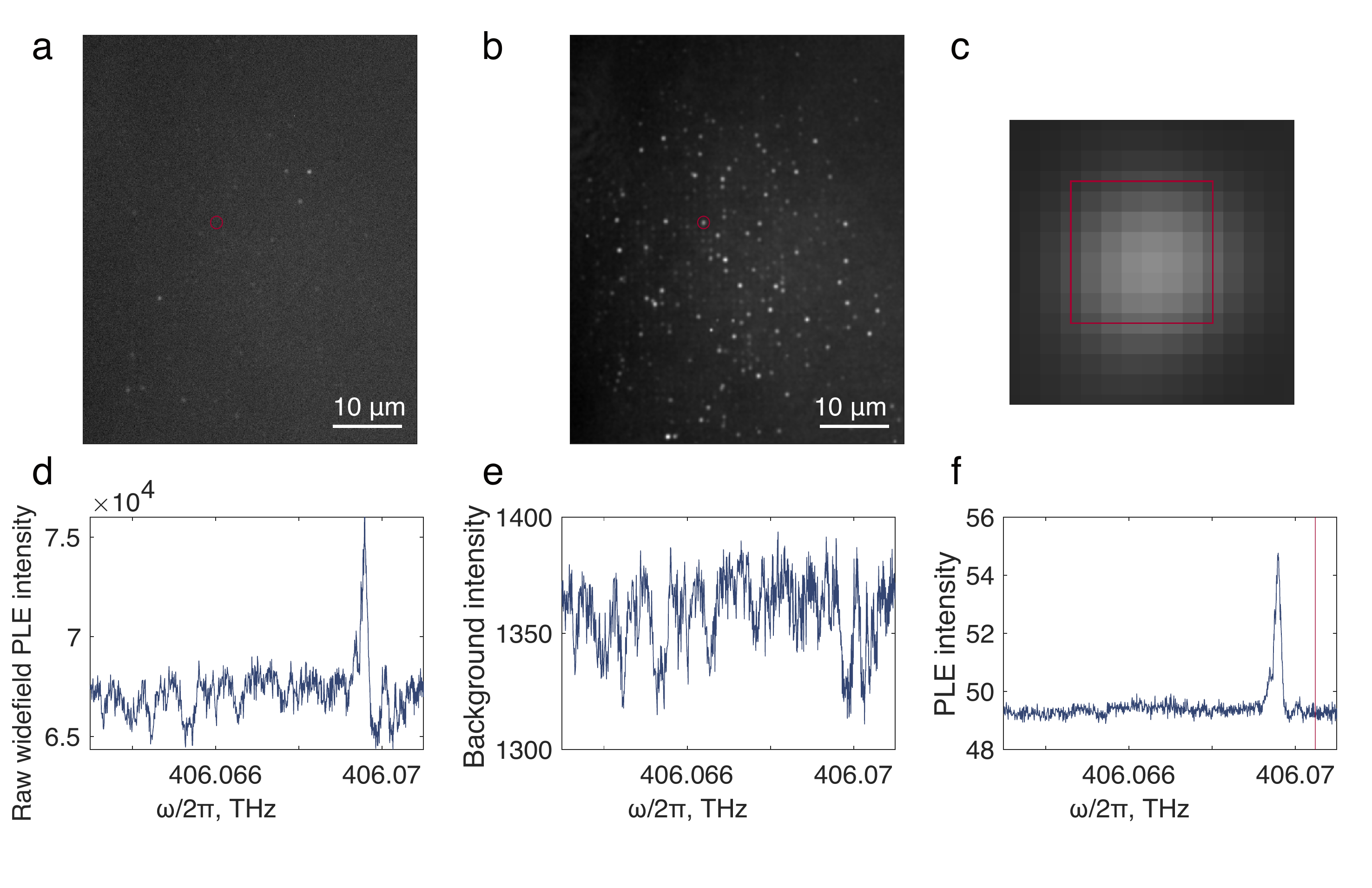}
\caption{\textbf{Analysis sequence to extract photoluminescence excitation (PLE) spectra from widefield fluorescence imaging.} \textbf{a,} Widefield image at a single frequency, 406.070480(2) THz, exposed for 1~s. A handful of SiVs are observed as diffraction-limited bright spots. \textbf{b,} Reconstructed image of all sites with widefield PLE peaks between 406.062985(2) and 406.070993(2) THz. For each spatial pixel, the value shown here represents the maximum pixel brightness for that pixel over all frequencies in the frequency interval A 2D Gaussian smoothing function was then applied to the resulting image. \textbf{c,} Zoomed-in view of the reconstructed site highlighted in \textbf{b}. All of the pixels that fall within the bounds of the red box are summed at each frequency slice and used to represent the widefield PLE intensity at that frequency. \textbf{d,} Raw widefield PLE data as a function of frequency for the site highlighted in \textbf{a}-\textbf{c}. \textbf{e,} Background spectrum constructed by averaging the pixel values of the entire widefield image at each frequency. \textbf{f,} Processed widefield PLE spectrum determined by dividing the raw spectrum by the background. The frequency at which the image shown in \textbf{a} was excited is indicated with a red line. }
\label{fig:widefield}
\end{figure}

Here, we present an example of the image processing performed to extract widefield PLE data from the raw widefield images collected. At each frequency step, the raw image was first cropped to remove area outside the QR region, as shown in Fig.~\ref{fig:widefield}a. Then, we filtered the image with a 2D Gaussian filter. We generated a 2D Gaussian kernel $f_g(x,y)$ with standard deviation $\sigma = 2$~pixels selected to approximately match the diffraction limited spot size:
\begin{equation}
f_g(x,y) = e^{-\frac{(x^2+y^2)}{2\sigma^2}}
\end{equation}
At each frequency step $\nu=\omega/2\pi$, we discretized and convolved the kernel with the image $M_\nu (x,y)$ via discrete 2-dimensional convolution to yield a filtered image $C_\nu (x,y)$:
\begin{equation}
    C_\nu(x,y) = \sum_i \sum_j M_\nu(i,j) f_g(x-i+1,y-j+1)
\end{equation}
Here, the range of $i$ and $j$ in the sum spans the dimension of the image and the kernel.

After filtering each image, we sectioned the data into frequency bands to facilitate image processing. At each frequency band spanning approximately 10~GHz, we generated an aggregate image $A(x,y)$ summarizing the fluorescence over the full band by taking the maximum value of each pixel over the images taken at all frequency steps. An example of an aggregate image generated for images taken while scanning the resonant laser from 405.8816-404.9005~THz is shown in Fig.~\ref{fig:widefield}b. We then used this aggregate image to identify candidate emitter sites for each frequency section, using the bandpass filtering and 2D Gaussian fitting process described previously in Methods Section~\ref{methods:confocal}, Eqs.~\ref{eq:lowpass}-\ref{eq:bandpassfilteredim}.  An example candidate site is circled in red in Fig.~\ref{fig:widefield}a-b. 

At each candidate emitter site, we determined the fluorescence intensity as a function of excitation frequency by binning pixels around the spatial center of the 2D Gaussian fit. An example site is shown in Fig.~\ref{fig:widefield}c, with the group of binned pixels shown in a red box. Then, this group of pixels was summed at each frequency step to find the fluorescence intensity as a function of excitation frequency, $I_{p,q}(\nu)$ for an emitter located at location $x=p$, $y=q$ in the convolved image $C_\nu (x,y)$:
\begin{equation}
    I_{p,q}(\nu) = \sum_{i=p-3}^{p+3} \sum_{j=q-3}^{q+3} C_\nu (i,j) 
    \label{eq:raw}
\end{equation}
We chose the number of pixels to include in the binned region to maximize the signal-to-noise ratio. Although we used the bandpass-filtered image $F(x,y)$ to generate the candidate sites, we generated the PLE intensity function $I_{p,q}(\nu)$ using the convolved images at each frequency step, $C_\nu (x,y)$. An example intensity function  $I_{p,q}(\nu)$ for the candidate SiV site is shown in Fig.~\ref{fig:widefield}d. In principle, both the emitter fluorescence signal $S_{p,q}$ at a site and background fluorescence from the sample $B$ both depend on the intensity of the excitation laser. A simple model is considered to account for fluctuating laser intensity $L(\nu)$, where $\nu$ is the frequency bin:
 \begin{equation}
     I_{p,q}(\nu) = S_{p,q}(\nu)L(\nu) + B L(\nu)
 \end{equation}
We model this as a linear process since the emitters are driven below saturation. We determined the intensity function $L(\nu)$ by calculating the average value of all pixels in the convolved image $C_\nu (x,y)$ at each frequency: 
\begin{equation}
    L(\nu) = \frac{1}{xy}\sum_x \sum_y C_\nu(x,y)
\end{equation}
While $L(\nu)$ accounts for global fluctuations of intensity, it does not account for regional fluctuations, including the point spread function of the widefield microscope. Finally, we determined the frequency-dependent SiV fluorescence signal at a given site $S_{p,q}(\nu)$ by dividing the measured intensity by the noise function:
\begin{equation}
S_{p,q}(\nu) + B= \frac{I_{p,q}(\nu)}{L(\nu)}
\end{equation}
An example of the frequency-dependent intensity function $L(\nu)$ at one site is shown Fig.
~\ref{fig:widefield}e, and the final corrected signal to be fit is shown in Fig.~\ref{fig:widefield}f. This model assumes that there are many color centers such that the emission frequencies of the population of color centers are uncorrelated, and that the dominant intensity fluctuation of the entire image is correlated with the excitation frequency. 

This assumption is valid for the case of Sample B, which has a broad inhomogeneous distribution of emitters even at individual sites. However, this assumption breaks down for Sample A, since the inhomogeneous distribution is narrow. Instead, for Sample A, we use a liquid-crystal noise-eater to stabilize the laser power, and fit the raw PLE data as given by Eq.~\ref{eq:raw}. \\
\\

\subsection{QR encoding and convolutional QR detection}\label{methods:qr}
Custom-designed binary codes etched into the diamond surface serve as our QR-style codes. These markers consist of a 2-dimensional array of bits, with 16 bits to encode the location, 4 bits to encode a version, 1 bit as a constant pad, and 3 bits as an error-detecting checksum.
\begin{figure}[htbp]
\centering
\includegraphics[width=\linewidth]{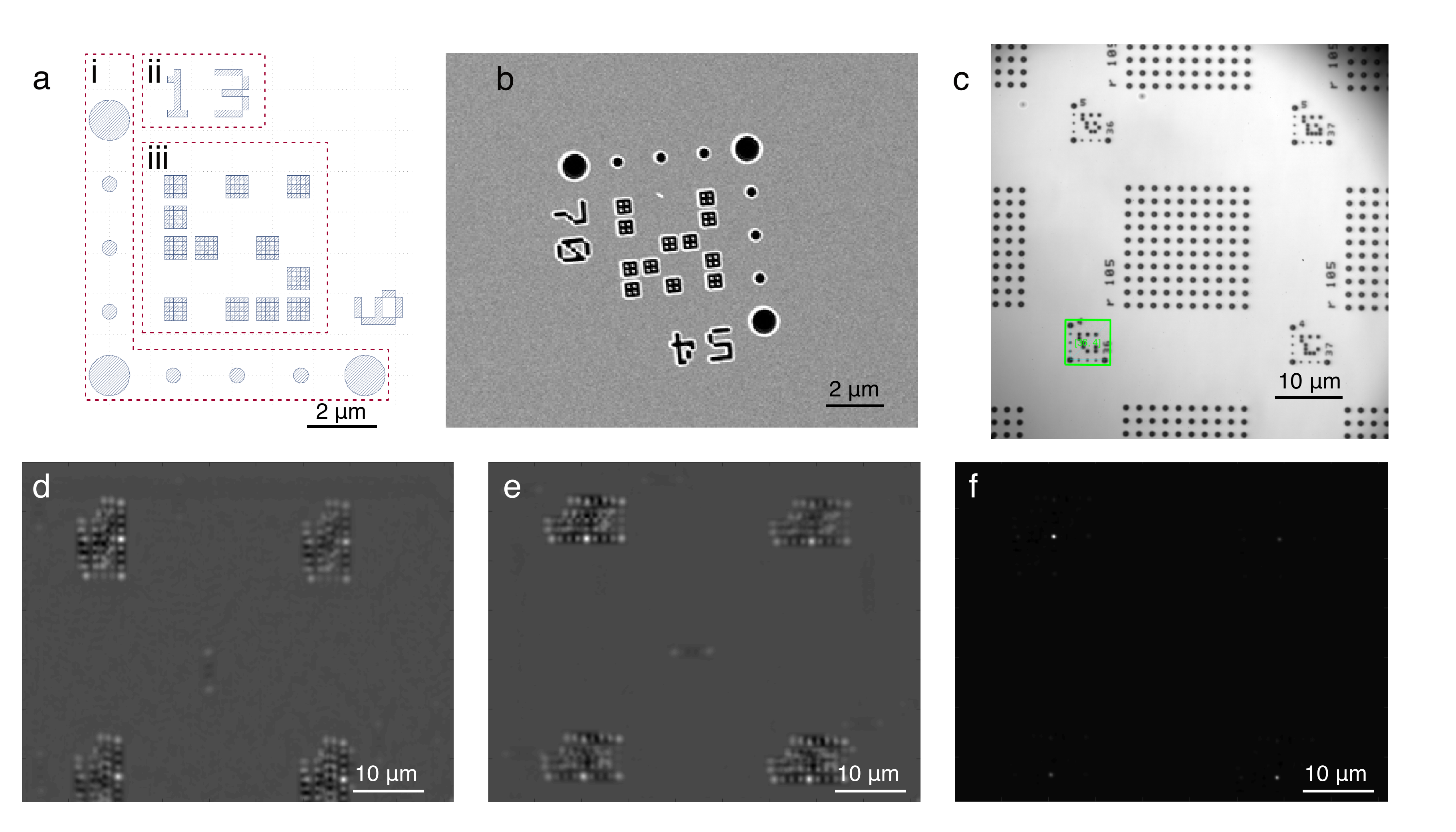}
\caption{\textbf{QR code design and image processing.} \textbf{a,} QR-style code design. \textbf{i,} Vertical and horizontal arrays of dots serve as alignment markers. \textbf{ii,} Row and column information is written for human-readability. \textbf{iii,} QR data: bits 2-5 encode the QR version, bits 7-14 encode the row, bits 15-22 encode the column, and bits 23-25 encode a checksum. \textbf{b,} SEM micrograph of a QR-style code fabricated in diamond. \textbf{c,} Widefield optical image showing the area bounded by four QR-style codes. \textbf{d,} Horizontal convolution to detect the horizontal QR arms. \textbf{e,} Vertical convolution to detect the vertical QR arms. \textbf{f,} Summing the cubes of the horizontal and vertical detections. The bright points that remain show the right angle between the QR alignment axes.}
\label{fig:qrs}
\end{figure}
Fig.~\ref{fig:qrs}\textbf{a}-\textbf{c} show an example of the layout and design of these markers, a SEM micrograph after the QR codes are patterned via lithography and etching, and a typical optical image under our microscopes.

We make use of a convolutional approach to quickly detect and decode the position of QR markers in an image. Given precharacterized sample rotation and scale, a kernel is generated, corresponding to the two large dots and three small dots of a single arm of the QR code bounding box (Fig.~\ref{fig:qrs}a). Convolving a flattened version of this image with this kernel and its 90$^\circ$ rotation yields two images with bright peaks corresponding to candidate positions for horizontal and vertical arms (Fig.~\ref{fig:qrs}d,e). The cubed sum of these two images leaves bright peaks at positions where horizontal and vertical arms are both detected, i.e. good candidates for full QR codes (Fig.~\ref{fig:qrs}f). At each candidate site, the $5\times5$ grid which stores the coded data is binarized and decoded, with errors detected by checksum failure. Altogether, this process is efficient at extracting QR candidates and data from images ($<100$~ms typical processing time), to the point that real-time video data is easily processed and displayed with detected QRs.

As an additional failsafe, we update a belief model for all QR candidates with positive checksum within a field of view via majority vote, to deduce the true location. This avoids situations where false-positive checksums yield failure. Importantly, this detection procedure is also robust when other features (e.g. fabricated structures) are present within the field of view. Altogether, this robust process for real-time QR detection enables rapid feedback to the user, along with automated traversal across a sample and reliable localization of fluorescent data with the global sample coordinate system.

\subsection{Advantage of widefield microscopy}\label{methods:speed}
Here, we present relevant considerations when choosing between confocal and widefield microscopy, and define the regimes in which our widefield approach results in a speed-up. The time to complete a photoluminescence excitation (PLE) experiment can vary widely depending on the specific approach: number of frequencies sampled, number of averages at each frequency point, whether a repump laser is used, microscope collection efficiency, and sensor efficiency all play a role. We therefore define a general scheme to determine whether a parallelized widefield approach will provide a time benefit.  The time to collect a confocal PLE scan at $m$ sites in a field of view, in $n$ frequency bins, is given by:

\begin{equation}
    \begin{split}
    t_{\text{confocal}} = m\big( t_{\text{move}} + t_{\text{tune laser, coarse}} +t_{\text{repump}} \\+ n(    t_{\text{tune laser, fine}} + t_{\text{collect,c}})\big)
    \end{split}
\end{equation}

Similarly, the time to collect a PLE scan in widefield for $m$ sites in a field of view is given by: 
\begin{equation}
    \begin{split}
    t_{\text{widefield}} =  t_{\text{tune laser, coarse}} +t_{\text{repump}}+ \\n(  t_{\text{tune laser, fine}} + t_{\text{collect,w}})
    \end{split}
\end{equation}
These equations hold in the case of a narrow inhomogeneous distribution: if all emitters are close in frequency, and $t_{\text{move}}$ is negligible, the time to collect data for $m$ emitters in a field of view in $n$ frequency bins scales simply as 
\begin{equation}
    \frac{t_{\text{confocal}}}{t_\text{widefield}} = m \frac{t_\text{collect, c}}{t_\text{collect, w}}
\end{equation}
As long as the integration time to collect the widefield data at each point is less than $m$ times the confocal case, our widefield approach is faster.

However, this approximation breaks down for the case of a large inhomogeneous distribution. Consider the far limiting case, in which there are $m$ emitters in the field of view that are all detuned in frequency such that they do not spectrally overlap. In this case, the time to collect data for $m$ emitters in a field of view in $n$ frequency bins is given by the ratio of collection times alone:
\begin{equation}
    \frac{t_{\text{confocal}}}{t_\text{widefield}} = \frac{t_\text{collect, c}}{t_\text{collect,w}}
\end{equation}
We therefore define an equation to determine the approximate speed-up factor (SF) of widefield over confocal PLE as a function of the number of color centers, $N$, the lifetime-limited linewidth $\gamma$, and the inhomogeneous distribution $\Gamma$: 
\begin{equation}
    \text{SF} = N \cdot \frac{\gamma}{\Gamma}\cdot\frac{t_\text{collect, c}}{t_\text{collect, w}}
\end{equation} 
A SF $> 1$ represents a speed-up for widefield over confocal microscopy. 

We now calculate the value of the SF for the two samples presented in the main text, Sample A and Sample B. Sample A represents the case of a very small inhomogeneous distribution. Four widefield scans were taken in the same field of view, covering the optical excitation frequencies around the A, B, C, and D transitions of the SiV in the absence of significant strain. All four transitions were found at 784 diffraction-limited widefield sites, representing 3,936 PLE peaks. We note that 200 of these diffraction-limited widefield sites contained eight peaks corresponding to two groups of SiV transition frequencies, but we count only unique spatial sites in our analysis, as the transition frequencies of the two SiV groups are so close in frequency. With $t_{\text{collect, c}} \approx t_{\text{collect, w}}$, and considering all emitters to lie in the same frequency bin, it would have taken $\text{SF} = 784$ times longer to measure the same strain distribution with our confocal microscope. For Sample B, with 40,186 peaks measured over 1.12~THz, we achieved a speed-up of $\text{SF} = 4$.

\subsection{Diamond fabrication}\label{methods:fab}
The fabrication procedure of patterning QR codes is the same as the one for fabricating nanopillar arrays. Each diamond sample was first deposited with silicon nitride (hard mask) via chemical vapor deposition. QR code and nanopillar array patterns were transferred into the nitride hard mask through electron beam lithography and CF$_4$ reactive-ion etching (RIE). Inductively-coupled plasma RIE with oxygen plasma subsequently transferred the mask patterns into bulk diamond. Finally, the samples were submerged in hydrofluoric acid to remove the nitride layer.

\subsection{Focused ion beam implantation}\label{methods:FIB}
We used a focused ion beam (FIB)~\cite{Schroder2017ScalableNanostructures} tool at the Ion Beam Laboratory (Sandia National Laboratories) to implant $^{29}$Si ions (spot size of about 50~nm $\times$ 45~nm) at an effective areal dose of $1\times10^0-1\times10^6$ ions/cm$^2$. The $^{29}$Si ion energy was 180~keV. After implantation, we annealed the sample at 1200$^\circ$C in an ultrahigh vacuum furnace. Finally, we tri-acid cleaned the sample in a 1:1:1 mixture of nitric, perchloric, and sulfuric acids by boiling for 1 hour.

\section*{Declarations}

\subsection*{Funding}
M.S. and E.B. acknowledge support from the NASA Space Technology Graduate Research Fellowship Program. I.C. acknowledges support from the National Defense Science and Engineering Graduate Fellowship Program, the National Science Foundation (NSF) EFRI ACQUIRE program EFMA-1641064, and NSF award DMR-1747426. M.S. and M.P.W. acknowledge partial support from the NSF Center for Integrated Quantum Materials (CIQM), DMR-1231319. K.C.C. acknowledges support from the NSF RAISE-TAQS CHE-1839155 and the MITRE Corporation Moonshot program. D.R.E. acknowledges partial support from the NSF Center for Quantum Networks (CQN), EEC-1941583. 

Distribution Statement A. Approved for public release. Distribution is unlimited. Diamond growth is based upon work supported by the National Reconnaissance Office (NRO) under Air Force Contract No. FA8702-15-D-0001. Any opinions, findings, conclusions or recommendations expressed in this material are those of the author(s) and do not necessarily reflect the views of the National Reconnaissance Office. 

The focused ion beam implantation work was performed at the Center for Integrated Nanotechnologies, an Office of Science User Facility operated for the US Department of Energy (DOE) Office of Science. Sandia National Laboratories is a multimission laboratory managed and operated by National Technology and Engineering Solutions of Sandia, LLC, a wholly owned subsidiary of Honeywell International, Inc., for the US DOE’s National Nuclear Security Administration under contract DE-NA-0003525. The views expressed in the article do not necessarily represent the views of the US DOE or the United States Government. This work made use of the Shared Experimental Facilities supported in part by the MRSEC Program of the NSF under award number DMR-1419807. 

\subsection*{Competing Interests}
The authors declare no competing interests.

\subsection*{Availability of data and materials}
Available on request to M.S.

\subsection*{Authors' contributions}
M.S., I.C., E.B., and M.P.W. conceived and performed experiments, developed software, and built optical setups. I.C. developed software and optical setups for widefield data collection and chip-scale sample screening. M.S. analyzed the data and wrote the manuscript with assistance from I.C., E.B., M.P.W., and K.C.C.  K.C.C. fabricated QR codes designed by M.P.W. SiV Sample A was produced by J.M., A.M., S.H., D.B., and P.B.D. SiV Sample B was focused ion beam implanted by M.T. and E.S.B. D.R.E. supervised the project. All authors discussed the results and contributed to the manuscript.
\bibliography{bib.bib}

\pagebreak
\newpage
\onecolumngrid

\section*{Extended data}
This work investigates approaches for large-scale data acquisition applied to fluorescent emitters.
We attach five Extended Data Videos which illustrate the scope of this work. 
These videos, also hosted on figshare at \url{https://figshare.com/s/185ccecd555aa21b5a50}, are condensed or cropped versions of larger scans to reduce file size to 80~MB. 

The full uncropped videos (with file size 940~MB) are hosted separately on figshare at \url{https://figshare.com/s/81a38527722002e996f6}. In the case of chip-scale scans, these show all of the fields of view investigated in each dataset to reduce selection bias.

Extended Data Video 1: \textbf{Widefield photoluminescence excitation (PLE) of silicon-vacancy centers in chemical-vapor deposition diamond.} \textbf{a, b, c, d,} Scanning over the A, B, C, and D optical transitions of the silicon-vacancy center in diamond, in a field of view. As expected, the C transition is the brightest of the four~\cite{Muller2014OpticalDiamondb}. The frequency of the resonant laser for a given frame is noted in the lower left of the panel. The scale bar represents 10 $\mu$m. Every frame represents approximately half a second of measurement  (video displayed at 10$\times$ speed). All panels are linearly scaled to the same fixed color limit.

Extended Data Video 2: \textbf{Chip-scale measurement of a sample with nano-fabricated pillars.} \textbf{a,} Bright-field images of each field of view, with boxes marking features which are detected as checksum-satisfying QR codes. Green boxes satisfy majority vote conditions for the field of view, and are considered correct. Yellow boxes usually result from noise which happens to satisfy the checksum, or from a bit error event which fails in majority vote condition. \textbf{b,} Fluorescence measurements with a 737~nm bandpass filter targeting SiV color centers. \textbf{c,} Fluorescence measurements with a 640~nm long-pass filter, with lower signal-to-noise ratio. This is likely due to fluorescence from NV centers or other emitters in the sample, especially in the bulk. Scale bars represent 10~$\mu$m. Every frame represents approximately 40 seconds of alignment and 20 seconds of measurement (video displayed at 600$\times$ speed). Each panel is linearly scaled to the maximum and minimum of each frame, except for \textbf{b}, which is scaled to a fixed color limit.

Extended Data Video 3: \textbf{Chip-scale measurement of a sample with deleterious noise.} \textbf{a,} Bright-field images with the same box color coding as Extended Data Video 2. This video demonstrates the robustness of our convolutional QR detection algorithm: we are able to detect QRs even through regions coated with stochastic particulates which pose a challenge for non-convolutional approaches. Every frame represents approximately 20 seconds of alignment and 20 seconds of measurement (video displayed at 400$\times$ speed). This panel is linearly scaled to the maximum and minimum of each frame.

Extended Data Video 4: \textbf{Chip-scale measurement of a sample with complex fabricated structures.} \textbf{a,} Bright-field images with the same box color coding as Extended Data Video 2. Despite the bright and varying nature of these fabricated structures, we are still able to conduct full-chip measurements. \textbf{b,} Fluorescence measurements at each field of view. Bright spots on the waveguides are attributed to either color centers, largely tin-vacancy in this case, or points where waveguide-coupled fluorescence is scattered upward. Scale bars represent 10 $\mu$m. Every frame represents approximately 20 seconds of alignment and 10 seconds of measurement (video displayed at 300$\times$ speed). \textbf{a} is linearly scaled to the maximum and minimum of each frame. \textbf{b} is scaled to a fixed color limit.

Extended Data Video 5: \textbf{Chip-scale measurement of sample B, a sample coimplanted with silicon, germanium, and tin.} \textbf{a,} Bright-field images with the same box color coding as Extended Data Video 2. \textbf{b,} Fluorescence images with a 640~nm long-pass filter. \textbf{c,} False-color images composited from three fluorescent measurements using 737~nm, 620~nm, and 600~nm bandpass filters with $\approx$10~nm bandwidth on the red, green, and blue color channels. These filters target the zero-phonon lines of silicon-, tin-, and germanium-vacancy color centers, respectively, though germanium appears as cyan due to its phonon sideband overlapping with the filter targeting the zero phonon line of tin. Scale bars represent 10 $\mu$m. Every frame represents approximately 30 seconds of alignment and 25 seconds of measurement (video displayed at 550$\times$ speed). \textbf{a} and \textbf{b} are linearly scaled to the maximum and minimum of each frame. \textbf{c} is shown with a fixed color range in log scale, to increase visibility of regions implanted with low dose.

\section*{Supplementary information}
\subsection*{S1. Cryogenic microscope}
\begin{figure}[htbp]
\centering
\includegraphics[width=.7\linewidth]{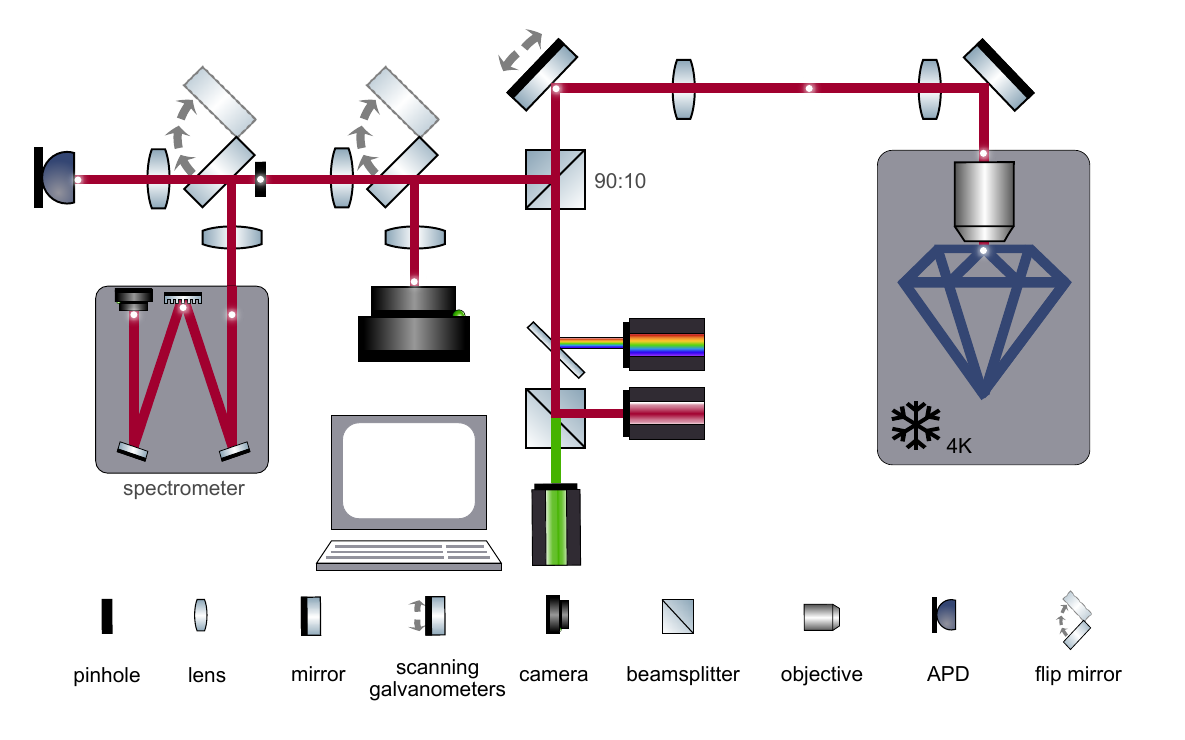}
\caption{\textbf{Cryogenic confocal microscope.} Details are provided in text.}
\label{fig:cf}
\end{figure}
We employ a homebuilt cryogenic confocal microscope as shown in Fig.~\ref{fig:cf}. Samples are cooled to $<4$~K in a Montana Instruments Cryostation equipped with three-axis piezoelectric stages (Attocube ANP101 Nanopositioners) to coarsely position the sample. Off-resonant excitation is provided by a 515~nm Hubner Cobolt laser. A Newport Velocity modulated by a free-space G\&H 3080 acousto-optic modulator is used to resonantly excite NVs, while an M$^2$ SolsTiS Ti:sapphire is used to resonantly excite SiVs. An incoherent white light source is also available for imaging. The excitation and collection paths are combined on a 90:10 beamsplitter, and scanning galvanometers enable beam steering over the field of view. Image planes and Fourier planes are indicated by white dots. 

Sample fluorescence is collected through the 90\% port in the beamsplitter, and the position of automated flip mirrors determine whether the fluorescence is collected on a camera (Photometrics Cascade 1K EMCCD), spectrometer (Princeton Instruments Isoplane), or avalanche photodiode (APD, Excelitas
SPCM-AQRH-14). For resonant excitation, the excitation laser frequency is blocked with an optical filter (longpass with 650~nm cutoff for NV experiments, 775/46~nm bandpass for SiV experiments)  such that only the phonon side-band reaches the APD. The full experimental sequence is controlled via a computer with homebuilt software.

\begin{figure}[htbp]
\centering
\includegraphics[width=.7\linewidth]{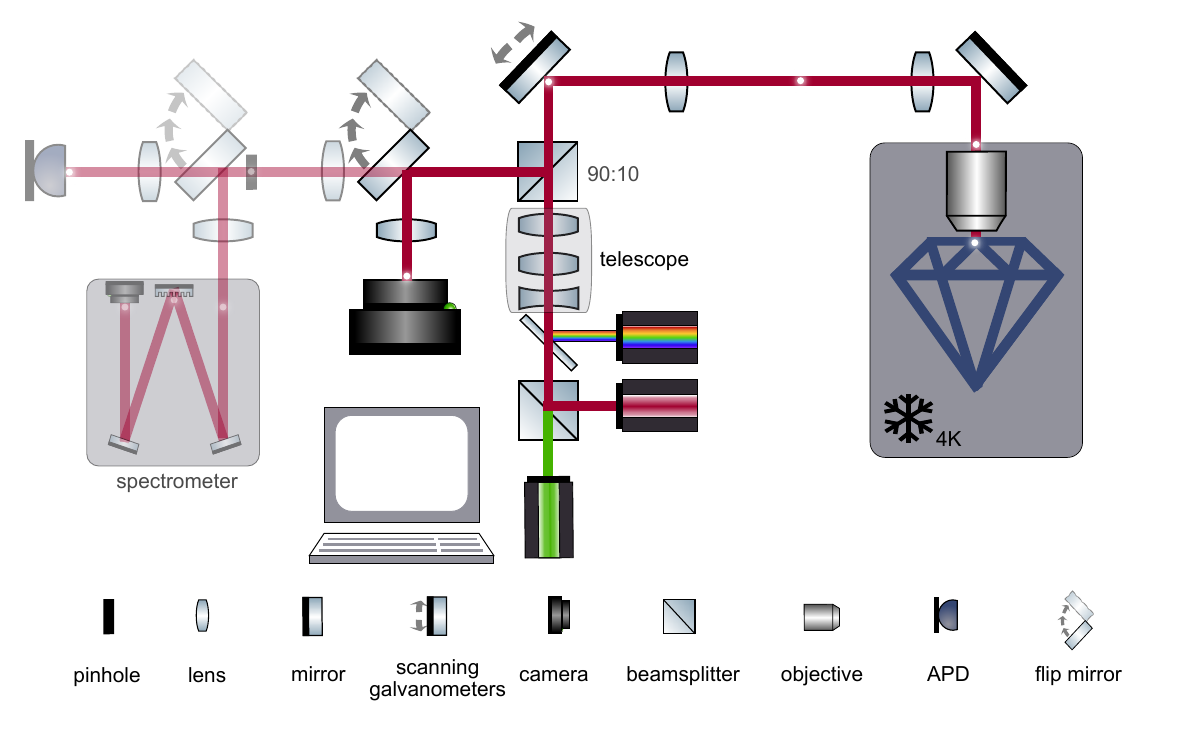}
\caption{\textbf{Cryogenic widefield microscope.} Details are provided in text.}
\label{fig:wf}
\end{figure}
Our confocal microscope is rapidly converted to a widefield microscope by adding a telescope -- consisting of a Galilean beam expander and a long focal-length lens -- the focus of which is mapped through the 4f to the back aperture of the objective, as shown in Fig.~\ref{fig:wf}. The telescope is placed using a magnetic mount.  The galvanometers are fixed in place for widefield measurements, and the fluorescence image is collected onto the EMCCD camera. 

\subsection*{S2. Room temperature microscope}
\begin{figure}[htbp]
\centering
\includegraphics[width=.55\linewidth]{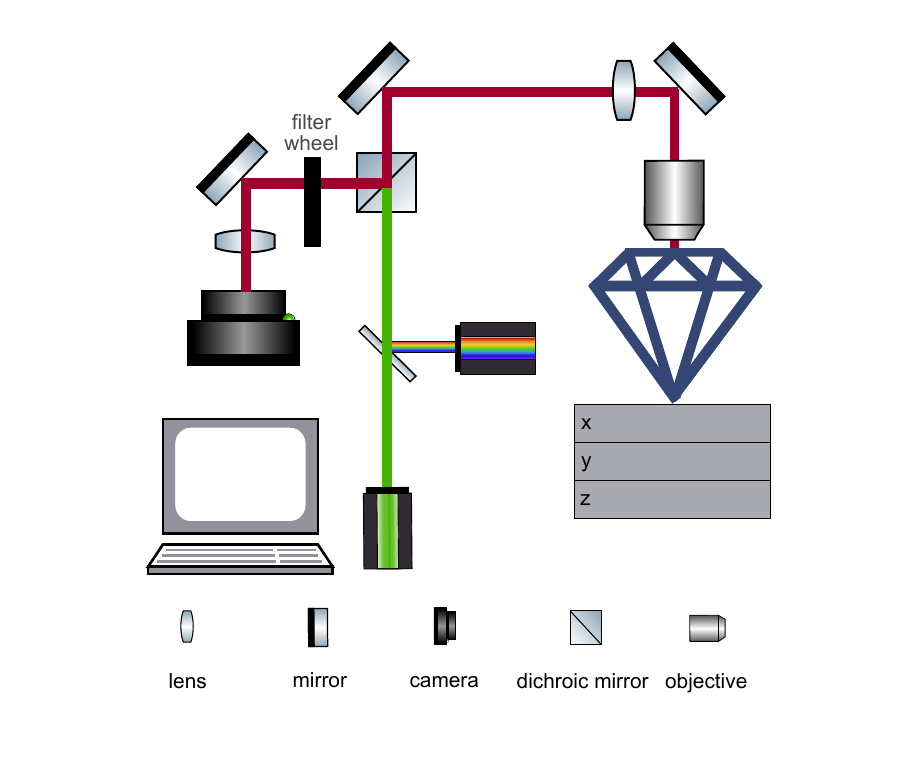}
\caption{\textbf{Room temperature widefield microscope.} Details are provided in text.}
\label{fig:romi}
\end{figure}

Our room temperature widefield microscope is shown in Fig.~\ref{fig:romi}. Off-resonant excitation is provided by a 532~nm Verdi laser. An incoherent white light source is also available for imaging. The excitation and collection paths are combined on a dichroic mirror, and fluorescence is collected on a Princeton Instruments ProEM EMCCD camera. An automated filter wheel enables selection of the fluorescence bandwidth to collect. A computer is used to feed back on Thorlabs 3-axis NanoMax linear translation stages to position the sample in X and Y, and to focus in Z, using custom image processing software and the fabricated QR codes. Examples are provided as Extended Data Videos. 

\subsection*{S3. Nitrogen-vacancy centers in diamond}
\begin{figure}[htbp]
\centering
\includegraphics[width=.6\linewidth]{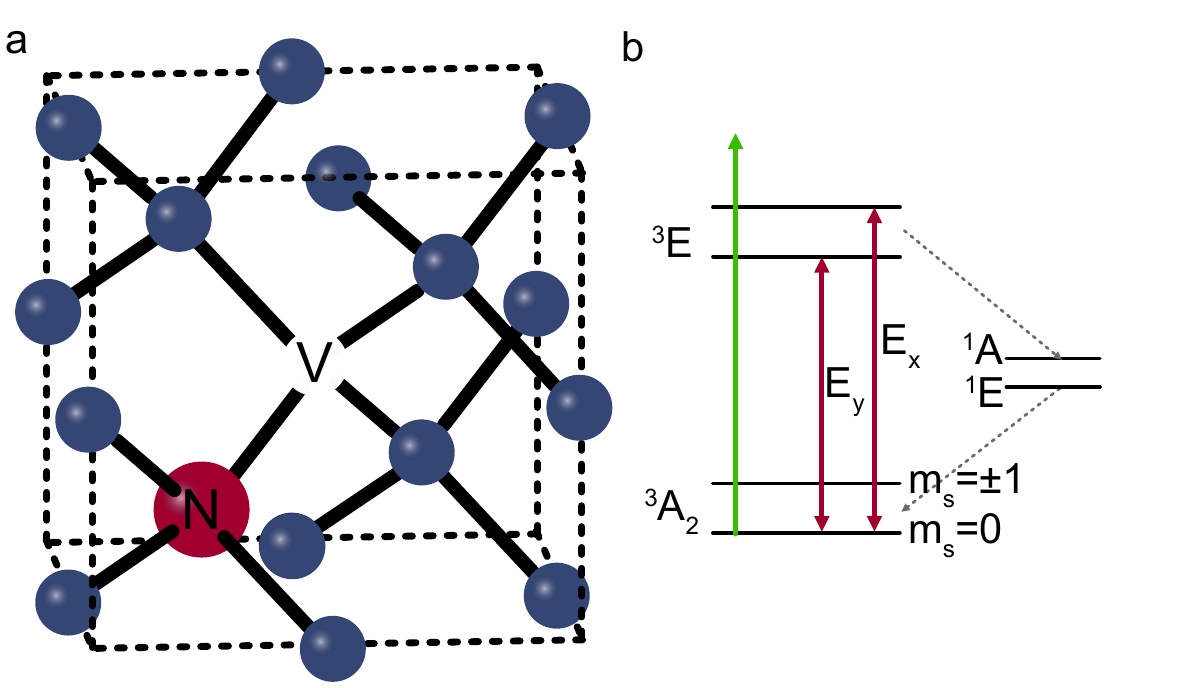}
\caption{\textbf{Nitrogen-vacancy center in diamond.} \textbf{a,} Ball and stick diagram showing one possible orientation of a nitrogen-vacancy center in diamond. Blue atoms represent carbon; the red atom is a substitutional nitrogen and the white V represents a carbon vacancy. There are four energetically equivalent NV orientations along each of the $\langle 111 \rangle$ family of crystallographic directions. \textbf{b,} Energy level diagram of a single negatively charged nitrogen-vacancy center in diamond under zero magnetic field. There are two relevant optical transitions between the excited state and ground state manifolds corresponding to $m_s=0$ transitions, labeled $E_x$ and $E_y$. The optically-active singlet pathway is labeled to the side. }
\label{fig:nv}
\end{figure}

The energy structure of a negatively charged nitrogen-vacancy center in diamond yields two optical transitions between the excited state and ground state manifolds that correspond to $m_s=0$ transitions, labeled $E_x$ and $E_y$. A schematic of a single nitrogen-vacancy center in diamond as well as an energy level structure showing the relevant optical transitions absent a magnetic field are shown in Fig.~\ref{fig:nv}. Strain breaks the degeneracy of $m_s=0$ optical transition, yielding two optical transitions distinct in frequency, $E_x$ and $E_y$. 
The response of NV centers to electric fields and strain can be equivalently described in the orbital basis $\{\ket{E_x},\ket{E_y}\}$ by the DC Stark perturbation~\cite{Bassett2011ElectricalFields}. The energy eigenvalues under the Stark effect are of the form 
\begin{equation}
    E_{\pm} = \hbar\bar{\omega} \pm h\frac{\delta}{2}
\end{equation}
for mean transition frequency $\bar{\omega}/2\pi$ and splitting $\delta$. Because $\bar{\omega}/2\pi$ scales with the longitudinal component and $\delta$ with the transverse component of the perturbing electric field, these eigenvalues enable observation of the local strain environment around a given NV center. 

In Fig.~2c of the main text, we observe that the change in mean frequency $\delta\bar{\omega}/2\pi$ splits into two groups following tri-acid cleaning. The dipole orientation of the NV is constrained to 8 possible configurations given by the $\langle 111 \rangle$ family of crystallographic directions and the dipole orientation of the nitrogen atom and the vacancy center.  

\subsection*{S4. Silicon-vacancy centers in diamond}
\begin{figure}[htbp]
\centering
\includegraphics[width=.6\linewidth]{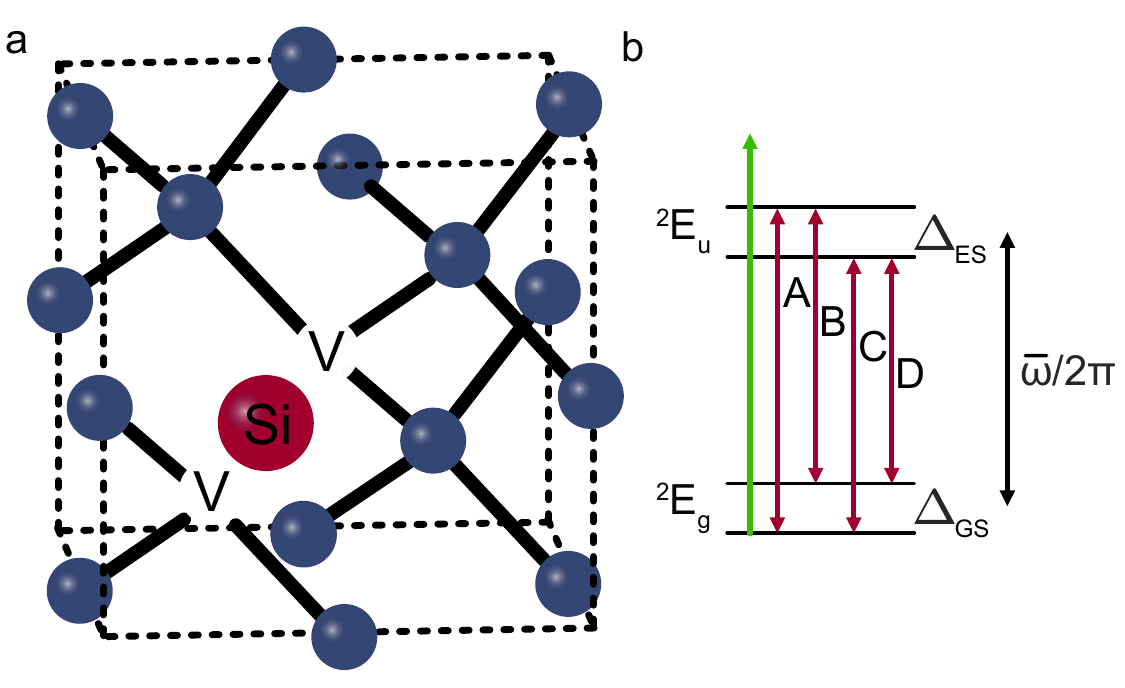}
\caption{\textbf{Silicon-vacancy center in diamond.} \textbf{a,} Ball and stick diagram showing one possible orientation of a silicon-vacancy center in diamond. Blue atoms represent carbon; the red atom is a substitutional silicon surrounded by a single split carbon vacancy. There are four energetically equivalent SiV orientations along each of the $\langle 111 \rangle$ family of crystallographic directions. \textbf{b,} Energy level diagram of a single negatively charged silicon-vacancy center in diamond under zero magnetic field. There are four relevant optical transitions between the excited state and ground state manifolds, labeled A, B, C, and D. Spin orbit coupling and strain split the ground state manifold by $\Delta_{GS}$ and the excited state manifold by $\Delta_{ES}$.}
\label{fig:siv}
\end{figure}

The energy structure of a negatively charged silicon-vacancy center in diamond yields four  optical transitions labeled A, B, C, and D, where A is the highest energy transition between the upper excited state and the lower ground state, and D is the lowest energy transition between the lower excited state and upper ground state. A schematic of a single silicon-vacancy center in diamond as well as an energy level structure showing the optical transitions under zero magnetic field are shown in Fig.~\ref{fig:siv}.

We label the SiV transitions by mean frequency $\bar{\omega}$, ground state splitting $\Delta_{\text{GS}}$, and excited state splitting $\Delta_{\text{ES}}$, which are calculated from experimentally-measured spectra: 
\begin{equation}
    2\pi\bar{\omega}=\frac{1}{4}\left( \omega_A + \omega_B + \omega_C + \omega_D \right)
\end{equation}
\begin{equation}
   2\pi \Delta_{\text{GS}} = \omega_C - \omega_D
\end{equation}
\begin{equation}
   2\pi \Delta_{\text{ES}} = \omega_B - \omega_D
\end{equation}

The strain environment around the SiVs found in Sample A can be determined directly from the measured optical transitions by using the eigenvalues of the diagonalized strain Hamiltonian for the $D_{3d}$-symmetric SiV~\cite{Meesala2018StrainDiamond}. The strain dependence of the mean zero-phonon line (ZPL) emission $\Delta_{ZPL}$ and using the strain susceptibility in the ground and excited state manifolds described in~\cite{Meesala2018StrainDiamond}, the ground state splitting $\Delta_{GS}$, and the excited state splitting $\Delta_{ES}$ are:

\begin{equation}
    \begin{split}
    \bar{\omega} = \Delta_{ZPL,0}+ \left(t_{\|,e} - t_{\|,g}\right)\epsilon_{zz} +\left(t_{\perp,e} - t_{\perp,g}\right) \left(\epsilon_{xx} + \epsilon_{yy}\right)
    \end{split}
\end{equation}
\begin{equation}
\begin{split}
    \Delta_{GS} =\big(\lambda_{SO,g}^2 +4\left( d_g \left(\epsilon_{xx}-\epsilon_{yy}\right)+f_g\epsilon_{yz}^2 \right)^2+4 \left( -2d_g \epsilon_{xy}+f_g\epsilon_{zx}\right)^2 \big)^{1/2}
    \end{split}
\end{equation}
\begin{equation}
\begin{split}
     \Delta_{ES} =\big( \lambda_{SO,e}^2 +4\left( d_e \left(\epsilon_{xx}-\epsilon_{yy}\right)+f_e\epsilon_{yz}^2 \right)^2+4 \left( -2d_e \epsilon_{xy}+f_e\epsilon_{zx}\right)^2 \big)^{1/2}
     \end{split}
\end{equation}
for strain susceptibilities $t_{\|,g(e)}$, $t_{\perp,g(e)}$,  $d_{g(e)}$, $f_{g(e)}$ in the ground (excited) state manifold. Axial strain $\epsilon_{zz}$ primarily yields a shift in the mean ZPL frequency $\bar{\omega}$, while transverse strain primarily yields increased splitting $\Delta_{GS}$ and $\Delta_{ES}$. The two classes of mean ZPL frequency $\bar{\omega}/2\pi$ observed in our experiment are indicative of two classes of axial strain $\epsilon_{zz}$ in Sample A, as shown in Fig.~3b-d of the main text. Similarly, the single population of $\Delta_{GS}$ and $\Delta_{ES}$ shown in Fig.~3e-f of the main text indicate a uniform transverse strain environment in Sample A. 
\end{document}